\documentclass[aps,amsmath,amssymb,prb,reprint,superscriptaddress,longbibliography]{revtex4-2}
\usepackage{graphicx}
\usepackage{color}
\usepackage[normalem]{ulem}

\newcommand{\rutwo}{RuO$_2$}
\newcommand{\titwo}{TiO$_2$}

\newcommand{\rhoxy}{$\rho_{xy}$}

\newcommand{\muohmcm}[1]{\ensuremath{#1\,\mu\Omega\textnormal{cm}}}

\newcommand{\seebeck}[1]{\ensuremath{#1\,\mu\textnormal{V K}^{-1}}}

\begin{document}
\title{Multiband transport in \rutwo{}}

\author{Florent Pawula}
\affiliation{Laboratoire de Cristallographie et Sciences des Mat\'eriaux (CRISMAT), Normandie Universit\'e, UMR6508 CNRS, ENSICAEN, UNICAEN, 14000 Caen, France}
\affiliation{Nantes Universit\'e, CNRS, Institut des Mat\'eriaux de Nantes Jean Rouxel (IMN)-UMR 6502, F-44000 Nantes, France}

\author{Ali Fakih}
\affiliation{Laboratoire de Cristallographie et Sciences des Mat\'eriaux (CRISMAT), Normandie Universit\'e, UMR6508 CNRS, ENSICAEN, UNICAEN, 14000 Caen, France}

\author{Ramzy~Daou}
\affiliation{Laboratoire de Cristallographie et Sciences des Mat\'eriaux (CRISMAT), Normandie Universit\'e, UMR6508 CNRS, ENSICAEN, UNICAEN, 14000 Caen, France}

\author{Sylvie~H\'ebert}
\affiliation{Laboratoire de Cristallographie et Sciences des Mat\'eriaux (CRISMAT), Normandie Universit\'e, UMR6508 CNRS, ENSICAEN, UNICAEN, 14000 Caen, France}

\author{Natalia Mordvinova}
\affiliation{Laboratoire de Cristallographie et Sciences des Mat\'eriaux (CRISMAT), Normandie Universit\'e, UMR6508 CNRS, ENSICAEN, UNICAEN, 14000 Caen, France}

\author{Oleg Lebedev}
\affiliation{Laboratoire de Cristallographie et Sciences des Mat\'eriaux (CRISMAT), Normandie Universit\'e, UMR6508 CNRS, ENSICAEN, UNICAEN, 14000 Caen, France}

\author{Denis Pelloquin}
\affiliation{Laboratoire de Cristallographie et Sciences des Mat\'eriaux (CRISMAT), Normandie Universit\'e, UMR6508 CNRS, ENSICAEN, UNICAEN, 14000 Caen, France}

\author{Antoine Maignan}
\affiliation{Laboratoire de Cristallographie et Sciences des Mat\'eriaux (CRISMAT), Normandie Universit\'e, UMR6508 CNRS, ENSICAEN, UNICAEN, 14000 Caen, France}

\date{10 June 2024}

\begin{abstract}
We present electrical and thermal transport measurements in single crystals of the metallic oxide \rutwo{}. The resistivity and Seebeck coefficient measured up to 970K confirm the metallic nature of transport. Magnetoresistance and Hall effect measurements as a function of orientation can be most easily described by a multiband transport model. We find that the ordinary Hall effect dominates any anomalous Hall signal in single crystals.
\end{abstract}
\maketitle

Fundamental research on \rutwo{} began 60 years ago when it was identified as a highly metallic conducting oxide  \cite{schaeferZurChemiePlatinmetalle1963,rogersCrystalChemistryMetal1969,rydenElectricalTransportProperties1970}. Its chemical stability and straightforward synthesis meant that it quickly found application as a component of precision resistors and was also identified early on as a potential barrier material for use in semiconductor devices  \cite{wittmerBarrierLayersPrinciples1984}. In the last twenty years it has seen renewed interest as a catalyst \cite{overSurfaceChemistryRuthenium2012}, as well as possible applications as a lithium storage material  \cite{balayaFullyReversibleHomogeneous2003}.

Experimental and theoretical work in the last few years has shown that even such simple and well known materials can host exotic states of matter. 
\rutwo{} has emerged as a candidate material hosting altermagnetism, the state where collinear antiferromagnetic ordering also breaks time reversal symmetry \cite{smejkalCrystalTimereversalSymmetry2020} due to the different symmetries of the magnetic and crystal lattices. However, the magnetic ordering in this system has not been unambiguously observed. Neutron scattering measurements on single crystals detected a magnetic reflection that would normally be forbidden in the rutile structure, which vanished by around 1000K \cite{berlijnItinerantAntiferromagnetismMathrm2017}. Resonant X-ray scattering \cite{zhuAnomalousAntiferromagnetismMetallic2019} later made a similar observation in both crystals and thin films.

Anomalous properties which depend on time-reversal symmetry breaking have since been observed in thin films of \rutwo{}, including spin transport \cite{baiObservationSpinSplitting2022,boseTiltedSpinCurrent2022}, magnetic circular dichroism \cite{fedchenkoObservationTimereversalSymmetry2023} and the anomalous Hall effect (AHE) \cite{fengAnomalousHallEffect2022}. Spin-resolved photoemission \cite{linObservationGiantSpin2024} also finds the $d$-wave symmetry expected from the altermagnetic state. 

While there seems to be a critical mass of observations of altermagnetic effects, questions about some of the original observations of magnetism, particularly in bulk crystals, have arisen \cite{loveseyMagneticPropertiesMathrm2022, smolyanyukFragilityMagneticOrder2024}. Muon spectroscopy, normally very sensitive to local moments, found no evidence for magnetism in bulk \rutwo{} \cite{hiraishiNonmagneticGroundState2024}. The recent controversy is very well summarised in Ref.~\onlinecite{smolyanyukFragilityMagneticOrder2024} whose calculations advance the hypothesis that altermagnetism in \rutwo{} only arises when the stoichiometric material is doped with holes.

Remarkably, despite being so well known, there have been relatively few studies of bulk transport properties of \rutwo{} in applied magnetic fields. 
In this paper we present measurements of the electrical and thermal transport in single crystals of \rutwo{} up to 970K to address three recent observations. Firstly, the magnetic peak in neutron scattering on bulk crystals was observed to be suppressed at a temperature between 900 and 1000K \cite{berlijnItinerantAntiferromagnetismMathrm2017}. A second experimental result was the observation of a kink in the resistivity of thin films grown on a variety of different substrates at 400K \cite{fengAnomalousHallEffect2022}. Previous electrical transport measurements above 300K consisted of just six widely spaced points up to 1000K \cite{rydenElectricalTransportProperties1970, glassfordElectronTransportProperties1994}. Our high resolution measurements show no sign of a phase transition in this temperature range in either the resistivity or Seebeck coefficient.

Thirdly, an anomalous Hall conductivity was observed in \rutwo (110) films \cite{fengAnomalousHallEffect2022}. The signatures of the AHE in particular were difficult to extract as it is only predicted to be present when N\'eel vector can be moved away from the easy c-axis by the applied magnetic field. The AHE does not therefore manifest with hysteresis in \rutwo{}, which makes it indistinguishable from the ordinary Hall effect with respect to the applied magnetic field direction. We have measured the magnetoresistance and Hall effect of single crystal samples as a function of orientation and sample quality, but find that we cannot distinguish any anomalous contribution to the Hall effect in the presence of strong ordinary contributions.

Single crystals of \rutwo{} were synthesized by the vapor-transport technique in a multi-zone tube furnace with flowing O$_2$ \cite{huangGrowthCharacterizationRuO21982}. An alumina crucible containing 5g of polycrystalline commercial \rutwo{} powder (Chempur, 99.9\%) was introduced in the first third of the tube length. The O$_2$ flow was set to 2 cm$^3$min$^{-1}$ while the temperature was ramped up to 1350°C in the furnace center over 24h, then at 60cm$^3$ min$^{-1}$ for the next 12 days. The furnace was powered down, allowed to cool overnight, and finally the O$_2$ flow was turned off. As synthesized single crystals have an elongated shape, typically of 0.5 to 1 mm long.

The X-ray powder diffraction (XRPD) patterns were performed using a standard powder diffractometer X’PERT Pro PANalytical (Philips) with Cu K$\alpha$ radiation in $\theta - 2\theta$ mode at room temperature.  
Electronic diffraction (ED) patterns were recorded using two different transmission electron microscopes a FEI TECNAI 30UT (Cs = 0.7mm) working at 300kV and a JEOL ARM200 cold FEG double-corrected microscope.

Transport properties were measured in a Physical Properties Measurement System (PPMS) cryostat equipped with a 9T magnet in the temperature range 2-400K.
Two single crystals of approximate dimensions 0.2 $\times$ 0.1 $\times$ 1.0 mm$^3$ were selected for their different residual resistivity ratios (RRR) of 80 (sample 1) and 12 (sample 2). Four-probe electrical resistivity was measured using silver paste (Dupont 4929 or 6838) contacts. Thermal properties were studied under high vacuum using custom thermal transport pucks for the single crystals. At temperatures up to 300K the standard one-heater two-thermometer technique was used with fine-wire thermocouples attached directly to the sample to detect the thermal gradient. Resistivity and Seebeck measurements were extended to 970K using miniature heaters and by thermally isolating the samples, still within the PPMS high vacuum environment.The RRR of sample 1 was checked after heating to 720K but did not change.

\begin{figure}
\includegraphics[width=0.45\textwidth]{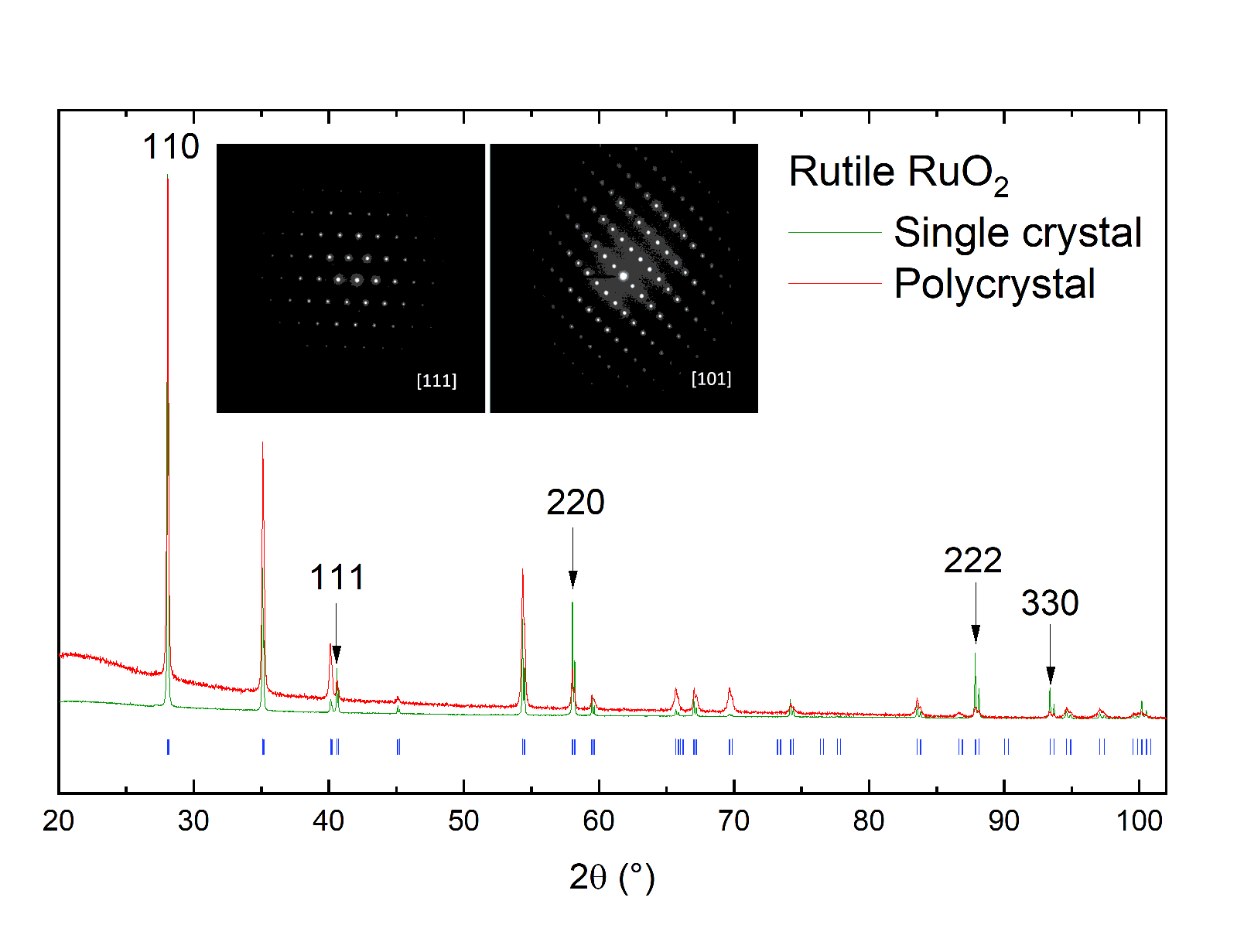}
\caption{Xray powder diffraction of \rutwo{} commercial powder starting material shows no difference to the ground single crystal. Inset: Single crystal ED patterns of \rutwo{} [111] (left) and [101] (right).}
\label{fig:xrpded}
\end{figure}

The XRPD patterns of the starting commercial powder and ground single crystals are similar (Fig.~\ref{fig:xrpded}). The single crystals are single phase rutile \rutwo{}. ED pattern analysis confirms the space group and lattice parameters along the [111] and the [101] directions, as shown in the inset to Fig.~\ref{fig:xrpded}.
The calculated lattice parameters from Lebail refinements, based on a tetragonal model with space group \#136 P42/mnm, are $a = 4.4908(2)\AA$ and $c = 3.1063(1)\AA$ for the single crystals, which is in excellent agreement with both early \cite{raoXrayStudiesThermal1969,rogersCrystalChemistryMetal1969,butlerCrystalGrowthElectrical1971} and recent reports \cite{berlijnItinerantAntiferromagnetismMathrm2017}. 
We note here that the lattice parameters of \titwo{}, the substrate used for epitaxial growth of thin films of \rutwo{} in several recent studies \cite{uchidaSuperconductivityUniquelyStrained2020,cuiInplaneHallEffect2023,fedchenkoObservationTimereversalSymmetry2023,fengAnomalousHallEffect2022,boseTiltedSpinCurrent2022} are $a = 4.594\AA$ and $c = 2.969\AA$, which results in considerable strain on the films with a strong dependence on orientation \cite{uchidaSuperconductivityUniquelyStrained2020}.

\begin{figure}
\includegraphics[width=0.5\textwidth]{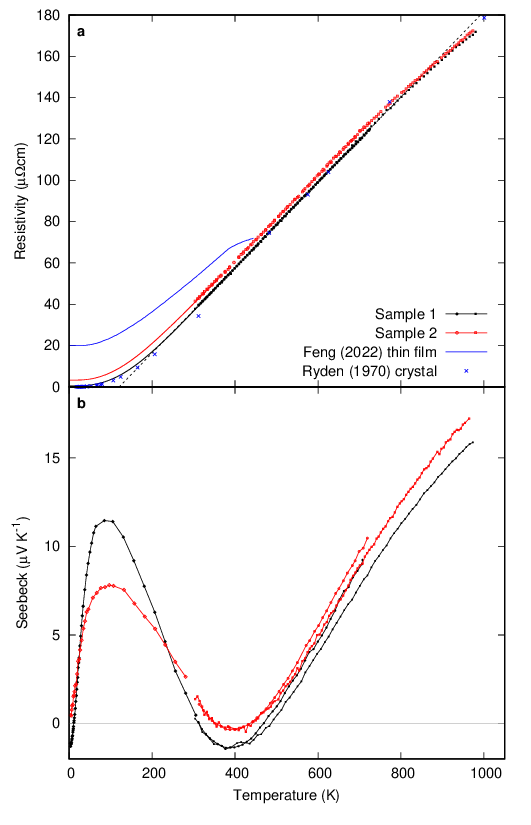}
\caption{a) Resistivity of single crystals of \rutwo{} with RRR 80 (sample 1) and RRR 12 (sample 2). The solid lines are data for the low temperature range, the different symbols refer to two separate high temperature runs on each sample (up to 720 and 970K) to confirm reproducibility. The resistivity of sample 2 has been slightly scaled to match the slope of that of sample 1 at 300K, but this is within the experimental uncertainty due to the size of the contacts. The dashed line is a linear fit to the data of sample 1 in the range 200-800K which emphasises the slight deviation from linearity at higher temperature. The resistivity of a thin film grown on \titwo{} is shown for comparison (blue curve from Ref.~\onlinecite{fengAnomalousHallEffect2022}). No sign of a feature like that seen near 380K in the blue curve is seen in the single crystal data. b) Seebeck coefficient in single crystals of \rutwo{}. The large positive peak in $S$ near 100K  and the small negative peak around 400K are slightly reduced as sample quality decreases. As the different symbols show, there is a slight difference between the high temperature runs that becomes more pronounced at higher temperatures, however the trend remains consistent.}
\label{fig:rho}
\end{figure}

The temperature dependent resistivity of two single crystals are shown in Fig.~\ref{fig:rho}a. The single crystals have resistivity of around \muohmcm{30} at 300K which is in reasonably good agreement with previous results \cite{rydenElectricalTransportProperties1970,linLowTemperatureElectrical2004}, given the geometric uncertainty involved in measuring finite-sized contacts on sub-millimeter samples. The residual resistivity ratios (RRR), $\rho(300K)/\rho(2K)$, are 80 for sample 1 and 12 for sample 2. This is lower than some previous reports \cite{rogersCrystalChemistryMetal1969, hiraishiNonmagneticGroundState2024}, but it is notable that the temperature dependence of the resistivity is largely similar over a wide range of sample quality. To a first approximation it appears that the impurity scattering does not have a temperature dependent component. The same appears to be true for thin film data, at least for temperatures below 380K, reproduced from Ref.~\onlinecite{fengAnomalousHallEffect2022} in Fig.~\ref{fig:rho}a where dislocations as well as impurities lead to residual resistivities that are generally higher than in crystals.

Resistivity data on these samples was taken up to 970K and is consistent with previous single crystal data, although only a few data points have been reported before in the range 300-1000K \cite{rydenElectricalTransportProperties1970}. Between around 300 and 800K, $\rho$ is approximately linear in temperature for both samples (see the dashed black line in Fig.~\ref{fig:rho}a). Above 800K there is a small downwards deviation from linearity. This contrasts with thin films grown on three different substrates, which all show a change in slope of the resistivity at around 380K, which was interpreted as a sign of the N\'eel temperature \cite{fengAnomalousHallEffect2022}.

Neutron scattering measurements indicate that the structural distortion associated with antiferromagnetism persists up to least 900K in bulk crystals \cite{berlijnItinerantAntiferromagnetismMathrm2017}, albeit with a short coherence length, while resonant X-ray scattering shows that this order persists with a coherence length of 4000\AA{}  up to 400K in single crystals (and is also present in thin films) \cite{zhuAnomalousAntiferromagnetismMetallic2019}. Photoemission has also confirmed the d-wave symmetry of the electronic structure expected for the collinear antiferromagnetic state \cite{linObservationGiantSpin2024}. However recent $\mu$SR measurements do not detect any magnetism in high-quality single crystals \cite{hiraishiNonmagneticGroundState2024}.

Up to around 800K, the resistivity curves are compatible with previous models \cite{glassfordElectronTransportProperties1994} that include Bloch-Grun\"{e}isen-type scattering from both acoustic and optical phonons, as well as an electron-electron scattering contribution, although our data are considerably more complete than the previously available set. The transport Debye and Einstein temperatures that we extract are around 500K and 900K. Beyond 800K there is a small downward deviation of the data from the best-fit model. Excluding the electron-electron scattering term results in a better fit over the whole temperature range, but does not change the extracted Debye and Einstein temperatures significantly.

The slope of extended region where the resistivity is linear in temperature can be used, in conjunction with the plasma frequency of 3.3eV \cite{glassfordElectronTransportProperties1994}, to estimate the transport scattering rate in terms of the so-called Planckian scattering rate. We find that $\frac{1}{\tau} \approx 3.5\frac{k_B T}{\hbar}$,  close to the range of values obtained for other materials where T-linear resistivity is observed. The Planckian scattering rate is a hypothetical upper bound on scattering derived from the smallest relevant timescale $\frac{\hbar}{k_B T}$, but it is not currently clear why so many materials follow this limit to within a small prefactor \cite{hartnollColloquiumPlanckianDissipation2022}. While we may therefore conclude that \rutwo{} is also a likely candidate for Planckian-bounded resistivity, it is not straightforward to employ these data to rigorously test this, given the multiband nature of conductivity in \rutwo{} \cite{grissonnancheLinearinTemperatureResistivity2021}. There is also the matter of the slight downward deviation from linearity at the highest temperatures, which would imply a change in the Fermi surface if this limit were still obeyed.

We show the Seebeck coefficient, $S$, of both samples in Fig.~\ref{fig:rho}b. We observe a peak in $S$ around 100K to a modest value of around \seebeck{10}. The size of the peak is reduced somewhat as the sample RRR decreases. A small negative contribution to $S$ at low temperature appears to be reduced in magnitude similarly.

A sample-dependent peak in $S$ may originate from either drag-type phenomena, or from the presence of multiple bands. In the former case, normal scattering of the electrons from, for example, the bath of phonons, which conserves the total momentum, must significantly outweigh the combination of electron-phonon umklapp and electron-defect scattering, which dissipate the momentum of the electron system. The thermal gradient applied to the phonon system in this way `drags' the electron system, usually leading to an enhancement of $S$. While the normal/umklapp scattering ratio is more or less fixed by the material properties, defect scattering depends most strongly on sample quality. For phonons the effect should be greatest at some fraction of the Debye temperature, at the point where phonon wavevectors are still too short to have effective umklapp processes, but normal processes are numerous. We note however that the Fermi surface topology, with bands that overlap the Brillouin zone boundary, allows umklapp events with even very short phonon wavevectors.

Alternatively, the presence of magnetism implies the existence of the analagous magnetic process, magnon drag, which would be active at temperatures some fraction of the N\'eel temperature.

On the other hand the presence of multiple bands can also account for variations in $S$ as a function of temperature and sample quality. The importance of contributions to $S$ of either sign varies slightly differently from sample to sample as a function of temperature depending on the precise details of scattering. In this way $S$ would provide complementary information to the Hall coefficient, which also reflects the sign of the carriers weighted by their relative mobilities. In the case of $S$ the weighting factors are the energy derivatives of the carrier numbers and the mobilities. If a single type of scattering dominates transport, as expected in both the high and low temperature limits, the latter will be the same for all the bands. The former implicates the energy dependence of the density of states over a width $\sim k_BT$, and $S$ can therefore be expected to acquire some unusual temperature dependence in the presence of narrow bands. This is the more likely scenario in the absence of a Fermi surface topology which enhances phonon drag.

We have extended our measurements of the Seebeck coefficient up to 970K. The behaviour is anomalous in comparison to other high-conductivity ruthenate compounds, where $S$ often saturates at a value of around \seebeck{30} in this temperature range \cite{hebertThermopowerQuadruplePerovskite2015,pawulaThermoelectricPropertiesStandard2021,mravljeThermopowerEntropyLessons2016a}. The continued monotonic rise of $S$ from 300 to 970K is consistent with persistent metallicity and no change in degrees of freedom. We saw no clear feature that might be linked to the onset of antiferromagnetism. A slight difference in the Seebeck coefficient measurements when extending measurements to 970K may be the result of a temperature offset in one of the thermocouples, possibly due to a small residue of silver paint. This small difference does not affect our conclusions.

Electronic structure calculations in the paramagnetic state predict that \rutwo{} is a compensated metal dominated by one large electron and one large hole band \cite{yavorskyInitioCalculationFermi1996}. This is in reasonable agreement with quantum oscillation measurements on single crystals. One electronic structure calculation in a collinear antiferromagnetic state \cite{fedchenkoObservationTimereversalSymmetry2023} showed more complex spin- and direction-dependent Fermi surfaces, although for transport in the c-direction the effect of these should be averaged out. Where multiple bands are present, the Seebeck coefficient can be represented as a sum of contributions from each band weighted by their conductivities: $S = \frac{\sum_i \sigma_i S_i}{\sum_i \sigma_i}$. Since the mobilities of each band can have a different temperature- and impurity-dependence, this leads naturally to a total $S$ that has impurity-dependent features such as the low temperature peak that we observe. While a value of $S$ can be calculated directly from electronic structure using e.g. the BoltzTraP software \cite{madsenBoltzTraP2ProgramInterpolating2018}, this contains within it the poor assumption that the relaxation time is isotropic and common to all bands, and therefore cancels out of the calculation of $S$. The usual result is smooth monotonic $S(T)$ curves that do not reflect the underlying multiband nature of conduction. Likewise, the expectations for the Seebeck coefficient from a Planckian scattering model are difficult to quantify in a multiband system, as it is apparent that even in the case of single-band material the scattering asymmetry with respect to energy that determines $S$ may acquire a momentum dependence \cite{georgesSkewedNonFermiLiquids2021,gourgoutSeebeckCoefficientCuprate2022}.


Motivated by the compatibility of the temperature dependent resistivity and Seebeck data with a multiband model of conduction, we performed measurements of the magnetoresistance and Hall effect at low temperatures. These also serve to compare directly with thin film measurements, where the Hall signal is enhanced for one particular orientation, but otherwise isotropic \cite{fengAnomalousHallEffect2022}.

\begin{figure*}
\includegraphics[width=0.7\textwidth]{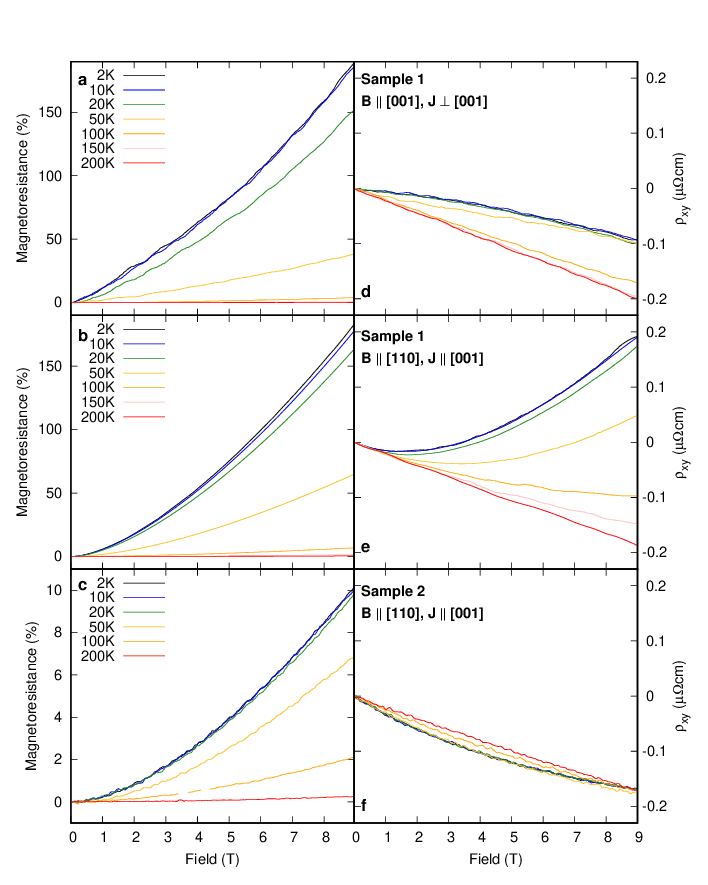}
\caption{The magnetoresistance (a-c) and Hall resistivity (d-f) of two samples of \rutwo{} in transverse magnetic fields of up to 9T. a,d) Sample 1, $RRR=80$, $B \parallel [001]$, $J \perp [001]$. b,e) Sample 1, $RRR=80$, $B \parallel [110]$, $J \parallel [001]$. c,f) Sample 2, $RRR=12$, $B \parallel [110]$, $J \parallel [001]$. The MR and nonlinearity of the Hall effect are strongly reduced when the RRR is reduced. The MR is unaffected by a change in the direction of measurement, but the nonlinearity of the Hall effect is strongly reduced.}
\label{fig:MR}
\end{figure*}

The transverse magnetoresistance (MR) and Hall effect of the two single crystals are presented in Fig.~\ref{fig:MR} in applied magnetic fields up to 9T, and for two different orientations. Sample 1 with RRR 80 was split into a long bar and a small cube. The cube was connected in a van der Pauw configuration with contacts elongated along the thickness to assure homogeneous current flow in the (001) plane. The magnetic field was applied along [001]. In this orientation (only measured for sample 1), we would not expect any AHE by symmetry\cite{fengAnomalousHallEffect2022}. The remaining part of sample 1 and sample 2 were oriented with their long axes, and current flow, along [001]. The magnetic field was applied close to the [110] direction in both cases.

Comparing Figs~\ref{fig:MR}a and \ref{fig:MR}b, we see that the MR is very similar for the two different orientations of the same crystal, reaching around 200\% in 9T. The MR is much greater for sample 1 than for sample 2 (Fig.\ref{fig:MR}c), which is typical of good metals as the RRR increases. Large MR can also arise easily in a compensated material like \rutwo{}.

The Hall resistivity at 200K is consistent across both samples and orientations (Figs\ref{fig:MR}d-f). It is linear in $B$ and has a value of around \muohmcm{-0.18} at 9T. This compares reasonably with the only other report of the magnitude of the Hall effect in single crystals from Ref.~\onlinecite{rydenElectricalTransportProperties1970}, from which we would expect a value of around \muohmcm{-0.10} at 300K. In comparison, the thin films where the AHE has been reported have $\rho_{xy}\sim$\muohmcm{-0.02}.

Comparing the Hall effect for the same orientation in different samples, $B\parallel [110], J \parallel [001]$ (Figs~\ref{fig:MR}e,f), we see a large positive contribution to $\rho_{xy}$ for sample 1 at higher fields and lower temperatures which is largely absent in lower-quality sample 2, where $\rho_{xy}$ remains negative and quasi-linear over all the field and temperature range. This is the orientation in which the AHE was observed in films as an excess negative  contribution to $\rho_{xy}$ visible only at lower temperatures \cite{fengAnomalousHallEffect2022}.

Finally, examining the Hall data for different orientations in sample 1, $B\parallel [001], J \perp [001]$ (Fig.~\ref{fig:MR}d) and $B\parallel [110], J \parallel [001]$ (Fig.~\ref{fig:MR}e), we find that the nonlinearity of the Hall effect is much reduced and there is no longer a sign change. The curvature of the Hall resistivity as a function of field also changes sense. The hole-like contribution to \rhoxy{} that is seen for $B\parallel [110]$ orientation  is no longer present in when $B \parallel [001]$. The sense of curvature indicates that it is an electron-like band that has the highest mobility in the latter case.

Our nonlinear MR and Hall data are again most naturally explained by a multiband scenario. A large, non-saturating MR is associated both with compensated systems and with open orbits. The latter have been identified in the [110] direction on the basis of angular dependent measurements \cite{marcusMeasurementMagnetoresistanceTransition1968}. A non-linear Hall effect is also expected in high-mobility compensated materials.

Electronic structure calculations of \rutwo{} in the paramagnetic state are dominated by two large bands that form the Fermi surface \cite{mattheissElectronicStructureRuO1976,fedchenkoObservationTimereversalSymmetry2023}. One calculation found additional small pockets \cite{yavorskyInitioCalculationFermi1996}, although quantum oscillation experiments are dominated by the larger bands with no conclusive identification of any frequencies with the smaller bands \cite{graebnerMagnetothermalOscillationsRuO1976}. Recent calculations in the collinear antiferromagnetic phase \cite{fedchenkoObservationTimereversalSymmetry2023} show that spin splitting leads to additional sheets which are larger than the small pockets previously suspected, however a full comparison to the older quantum oscillation data is lacking.

Anisotropy in the Hall effect is also implied by the tetragonal symmetry, where $\rho_{xy}$ is not required to be the same as $\rho_{xz}$. In this respect the observation of isotropic Hall resistivity for $B \parallel [001]$ and $B \parallel [100]$ in thin films was already quite remarkable \cite{fengAnomalousHallEffect2022}. In the case of a moderately ellipsoidal Fermi surface, such as the large electron sheet in the paramagnetic electronic structure of \rutwo{} \cite{yavorskyInitioCalculationFermi1996,fedchenkoObservationTimereversalSymmetry2023}, we may be confident that the sign (and approximate magnitude) of the ordinary Hall coefficient would remain the same when measured in these two orientations. For a multiply-connected Fermi surface sheet, such as the hole pocket in the same calculations, a more complex behaviour may arise as the orientation of the magnetic field is changed. Regions of different curvature can give rise to both hole- and electron-like trajectories on the same sheet, and therefore to opposite sign contributions to the Hall effect.

The electronic structure in the collinear antiferromagnetic state with spin-splitting is more complex \cite{fedchenkoObservationTimereversalSymmetry2023} but the qualitative topology of the Fermi surface is similar, with some surfaces that are self-contained and some that are connected across the Brillouin zone boundary. A strong difference between the ordinary contributions to $\rho_{xy}$ and $\rho_{xz}$ is therefore likely.

One potentially relevant observation was made in a systematic study of \rutwo{} films grown on all orientations of \titwo{} \cite{uchidaSuperconductivityUniquelyStrained2020}, where (110) oriented films were the only ones to show superconductivity at low temperatures. These films were also the only orientation to show anisotropy of the in-plane lattice parameter, where the $a_{110}$ and $a_{1\bar{1}0}$ lengths became different, corresponding to a lifting of the tetragonal symmetry. That this orientation should also be the only one to show AHE is an observation that motivates further investigation.

On the other hand, the Hall resistivity that we observe at 9T is an order of magnitude larger than the values observed in the thin films where the AHE was reported. While the slope of the resistivity in thin films is compatible with single crystals up to 380K, the change of slope seen there in films grown on many different substrates is a qualitative difference that implies some intrinisic difference arising either from the strained lattice or a different level of doping.

It is well known that oxygen pressure during growth controls the sample quality and can have a large effect on the Hall coefficient \cite{steevesELECTRONICTRANSPORTPROPERTIES}, with an $n$-to-$p$-type transition occuring with increasing oxygen deficiency. The small $n$-type Hall coefficient in thin films should be seen as a consequence of this effect. Another recent measurement on (101) thin films grown on \titwo{} showed an ordinary Hall contribution of around \muohmcm{0.05} in 2T \cite{cuiInplaneHallEffect2023}, which is closer to the expected bulk value. 

It is only when the magnetic field is oriented along [110] that an additional electron-like AHE is observed in films. This becomes smaller as the temperature is increased, which the authors of Ref.~\cite{fengAnomalousHallEffect2022} interpreted as the reduction in the magnitude of the N\'eel vector with increasing temperature. $\sigma_{AHE}$ vanishes around 200K while the N\'eel temperature, if the feature in resistivity is to be believed, is closer to 400K.

We would expect to see the AHE in all of our samples if they are indeed altermagnetic, and moreover we would expect it to persist to higher temperature if its presence depends on the N\'eel temperature, which is thought to be in excess of 900K in bulk material \cite{berlijnItinerantAntiferromagnetismMathrm2017}. It is not straightforward to compare the thin film and single crystal data, however. In our samples, the closest match to the sign and curvature of the AHE seen in films is for sample 2, where the Hall resistivity becomes more negative with decreasing temperature. However the comparison is still imperfect, as by 9T the low- and high-temperature \rhoxy{} curves converge again. An extended study showed that \rhoxy{} saturates above 60T in thin films \cite{tschirnerSaturationAnomalousHall2023}. Qualitatively then, the comparable features of the Hall effect in both thin films and crystals is an anisotropy with respect to the magnetic field direction, with a greater non-linearity for fields applied away from the c-axis. However, the magnitude of both the high- and low- temperature Hall signals, the sign of the low-temperature contribution, and the field-scale required are all different to those observed in thin films.

In summary, in single crystals we observe monotonically increasing resistivity and Seebeck coefficient at temperatures above 300K, with a small dependence on sample quality. At lower temperatures we see strong magnetoresistance that depends on RRR and temperature, but not on the orientation of the magnetic field. We see a non-linear Hall effect with a strong hole-like contribution at low temperatures for $B \parallel [110]$. This contribution is greatly reduced for $B \parallel [001]$. The Hall effect is also less field- and temperature-dependent when the RRR is reduced.

All of these features could be accounted for by multiband, compensated or nearly-compensated conduction. The increased RRR of our single crystals with respect to the thin films measured in Ref.~\onlinecite{fengAnomalousHallEffect2022} means that high-mobility carriers produce an ordinary Hall effect that is substantially non-linear at much lower magnetic fields. This signal may not therefore be reliably separated from the AHE, if it is indeed present. The magnitude of the intrinsic AHE arising from non-collinear magnetism observed in films is around 1000 $(\Omega cm)^{-1}$ at 50T \cite{fengAnomalousHallEffect2022}, while here $\sigma_{xy}$ is already $10^5 (\Omega cm)^{-1}$ in 9T. We note additionally that any defect-dependent sources of AHE (skew and side-jump scattering) should be further reduced as the RRR is increased. The resistivity anomaly in thin films at 380K is a further point of difference that remains to be explained. It is also noteworthy that the AHE is only observed in films where the in-plane rotational symmetry is lifted \cite{uchidaSuperconductivityUniquelyStrained2020}, when in principle it should be observed in other directions too, according to calculations \cite{fengAnomalousHallEffect2022}.

It is currently difficult to describe the wealth of new and old data on thin film and single crystals of \rutwo{} within a single consistent model. A plausible scenario is that a combination of oxygen deficiency (and possibly strain effects \cite{sasabeFerroicOrderAnisotropic2023}) may account for the diversity of current observations, although even if we limit ourselves to single crystal studies there is still fundamental disagreement between probes as to whether the altermagnetic state exists \cite{berlijnItinerantAntiferromagnetismMathrm2017,hiraishiNonmagneticGroundState2024}. A recent thorough reexamination of available data suggests that the altermagnetic state may only be stabilised under certain conditions of stoichiometry \cite{smolyanyukFragilityMagneticOrder2024}.

\begin{acknowledgments}
The authors would like to thank Alaska Subedi and Makariy Tanatar for helpful discussions.
\end{acknowledgments}


\begin{thebibliography}{39}%
\makeatletter
\providecommand \@ifxundefined [1]{%
 \@ifx{#1\undefined}
}%
\providecommand \@ifnum [1]{%
 \ifnum #1\expandafter \@firstoftwo
 \else \expandafter \@secondoftwo
 \fi
}%
\providecommand \@ifx [1]{%
 \ifx #1\expandafter \@firstoftwo
 \else \expandafter \@secondoftwo
 \fi
}%
\providecommand \natexlab [1]{#1}%
\providecommand \enquote  [1]{``#1''}%
\providecommand \bibnamefont  [1]{#1}%
\providecommand \bibfnamefont [1]{#1}%
\providecommand \citenamefont [1]{#1}%
\providecommand \href@noop [0]{\@secondoftwo}%
\providecommand \href [0]{\begingroup \@sanitize@url \@href}%
\providecommand \@href[1]{\@@startlink{#1}\@@href}%
\providecommand \@@href[1]{\endgroup#1\@@endlink}%
\providecommand \@sanitize@url [0]{\catcode `\\12\catcode `\$12\catcode
  `\&12\catcode `\#12\catcode `\^12\catcode `\_12\catcode `\%12\relax}%
\providecommand \@@startlink[1]{}%
\providecommand \@@endlink[0]{}%
\providecommand \url  [0]{\begingroup\@sanitize@url \@url }%
\providecommand \@url [1]{\endgroup\@href {#1}{\urlprefix }}%
\providecommand \urlprefix  [0]{URL }%
\providecommand \Eprint [0]{\href }%
\providecommand \doibase [0]{https://doi.org/}%
\providecommand \selectlanguage [0]{\@gobble}%
\providecommand \bibinfo  [0]{\@secondoftwo}%
\providecommand \bibfield  [0]{\@secondoftwo}%
\providecommand \translation [1]{[#1]}%
\providecommand \BibitemOpen [0]{}%
\providecommand \bibitemStop [0]{}%
\providecommand \bibitemNoStop [0]{.\EOS\space}%
\providecommand \EOS [0]{\spacefactor3000\relax}%
\providecommand \BibitemShut  [1]{\csname bibitem#1\endcsname}%
\let\auto@bib@innerbib\@empty
\bibitem [{\citenamefont {Sch{\"a}fer}\ \emph {et~al.}(1963)\citenamefont
  {Sch{\"a}fer}, \citenamefont {Schneidereit},\ and\ \citenamefont
  {Gerhardt}}]{schaeferZurChemiePlatinmetalle1963}%
  \BibitemOpen
  \bibfield  {author} {\bibinfo {author} {\bibfnamefont {H.}~\bibnamefont
  {Sch{\"a}fer}}, \bibinfo {author} {\bibfnamefont {G.}~\bibnamefont
  {Schneidereit}},\ and\ \bibinfo {author} {\bibfnamefont {W.}~\bibnamefont
  {Gerhardt}},\ }\bibfield  {title} {\bibinfo {title} {{Zur Chemie der
  Platinmetalle. {{RuO}}\textsubscript{2} Chemischer Transport, Eigenschaften,
  thermischer Zerfall}},\ }\href {https://doi.org/10.1002/zaac.19633190514}
  {\bibfield  {journal} {\bibinfo  {journal} {Zeitschrift f\"ur anorganische
  und allgemeine Chemie}\ }\textbf {\bibinfo {volume} {319}},\ \bibinfo {pages}
  {327} (\bibinfo {year} {1963})}\BibitemShut {NoStop}%
\bibitem [{\citenamefont {Rogers}\ \emph {et~al.}(1969)\citenamefont {Rogers},
  \citenamefont {Shannon}, \citenamefont {Sleight},\ and\ \citenamefont
  {Gillson}}]{rogersCrystalChemistryMetal1969}%
  \BibitemOpen
  \bibfield  {author} {\bibinfo {author} {\bibfnamefont {D.~B.}\ \bibnamefont
  {Rogers}}, \bibinfo {author} {\bibfnamefont {R.~D.}\ \bibnamefont {Shannon}},
  \bibinfo {author} {\bibfnamefont {A.~W.}\ \bibnamefont {Sleight}},\ and\
  \bibinfo {author} {\bibfnamefont {J.~L.}\ \bibnamefont {Gillson}},\
  }\bibfield  {title} {\bibinfo {title} {Crystal chemistry of metal dioxides
  with rutile-related structures},\ }\href
  {https://doi.org/10.1021/ic50074a029} {\bibfield  {journal} {\bibinfo
  {journal} {Inorganic Chemistry}\ }\textbf {\bibinfo {volume} {8}},\ \bibinfo
  {pages} {841} (\bibinfo {year} {1969})}\BibitemShut {NoStop}%
\bibitem [{\citenamefont {Ryden}\ \emph {et~al.}(1970)\citenamefont {Ryden},
  \citenamefont {Lawson},\ and\ \citenamefont
  {Sartain}}]{rydenElectricalTransportProperties1970}%
  \BibitemOpen
  \bibfield  {author} {\bibinfo {author} {\bibfnamefont {W.~D.}\ \bibnamefont
  {Ryden}}, \bibinfo {author} {\bibfnamefont {A.~W.}\ \bibnamefont {Lawson}},\
  and\ \bibinfo {author} {\bibfnamefont {C.~C.}\ \bibnamefont {Sartain}},\
  }\bibfield  {title} {\bibinfo {title} {Electrical {{Transport Properties}} of
  {{IrO}}\textsubscript{2} and {{RuO}}\textsubscript{2}},\ }\href
  {https://doi.org/10.1103/PhysRevB.1.1494} {\bibfield  {journal} {\bibinfo
  {journal} {Physical Review B}\ }\textbf {\bibinfo {volume} {1}},\ \bibinfo
  {pages} {1494} (\bibinfo {year} {1970})}\BibitemShut {NoStop}%
\bibitem [{\citenamefont {Wittmer}(1984)}]{wittmerBarrierLayersPrinciples1984}%
  \BibitemOpen
  \bibfield  {author} {\bibinfo {author} {\bibfnamefont {M.}~\bibnamefont
  {Wittmer}},\ }\bibfield  {title} {\bibinfo {title} {Barrier layers:
  {{Principles}} and applications in microelectronics},\ }\href
  {https://doi.org/10.1116/1.572580} {\bibfield  {journal} {\bibinfo  {journal}
  {Journal of Vacuum Science \& Technology A: Vacuum, Surfaces, and Films}\
  }\textbf {\bibinfo {volume} {2}},\ \bibinfo {pages} {273} (\bibinfo {year}
  {1984})}\BibitemShut {NoStop}%
\bibitem [{\citenamefont {Over}(2012)}]{overSurfaceChemistryRuthenium2012}%
  \BibitemOpen
  \bibfield  {author} {\bibinfo {author} {\bibfnamefont {H.}~\bibnamefont
  {Over}},\ }\bibfield  {title} {\bibinfo {title} {Surface {{Chemistry}} of
  {{Ruthenium Dioxide}} in {{Heterogeneous Catalysis}} and
  {{Electrocatalysis}}: {{From Fundamental}} to {{Applied Research}}},\ }\href
  {https://doi.org/10.1021/cr200247n} {\bibfield  {journal} {\bibinfo
  {journal} {Chemical Reviews}\ }\textbf {\bibinfo {volume} {112}},\ \bibinfo
  {pages} {3356} (\bibinfo {year} {2012})}\BibitemShut {NoStop}%
\bibitem [{\citenamefont {Balaya}\ \emph {et~al.}(2003)\citenamefont {Balaya},
  \citenamefont {Li}, \citenamefont {Kienle},\ and\ \citenamefont
  {Maier}}]{balayaFullyReversibleHomogeneous2003}%
  \BibitemOpen
  \bibfield  {author} {\bibinfo {author} {\bibfnamefont {P.}~\bibnamefont
  {Balaya}}, \bibinfo {author} {\bibfnamefont {H.}~\bibnamefont {Li}}, \bibinfo
  {author} {\bibfnamefont {L.}~\bibnamefont {Kienle}},\ and\ \bibinfo {author}
  {\bibfnamefont {J.}~\bibnamefont {Maier}},\ }\bibfield  {title} {\bibinfo
  {title} {Fully {{Reversible Homogeneous}} and {{Heterogeneous Li Storage}} in
  {{RuO}}\textsubscript{2} with {{High Capacity}}},\ }\href
  {https://doi.org/10.1002/adfm.200304406} {\bibfield  {journal} {\bibinfo
  {journal} {Advanced Functional Materials}\ }\textbf {\bibinfo {volume}
  {13}},\ \bibinfo {pages} {621} (\bibinfo {year} {2003})}\BibitemShut
  {NoStop}%
\bibitem [{\citenamefont {{\v S}mejkal}\ \emph {et~al.}(2020)\citenamefont {{\v
  S}mejkal}, \citenamefont {{Gonz{\'a}lez-Hern{\'a}ndez}}, \citenamefont
  {Jungwirth},\ and\ \citenamefont
  {Sinova}}]{smejkalCrystalTimereversalSymmetry2020}%
  \BibitemOpen
  \bibfield  {author} {\bibinfo {author} {\bibfnamefont {L.}~\bibnamefont {{\v
  S}mejkal}}, \bibinfo {author} {\bibfnamefont {R.}~\bibnamefont
  {{Gonz{\'a}lez-Hern{\'a}ndez}}}, \bibinfo {author} {\bibfnamefont
  {T.}~\bibnamefont {Jungwirth}},\ and\ \bibinfo {author} {\bibfnamefont
  {J.}~\bibnamefont {Sinova}},\ }\bibfield  {title} {\bibinfo {title} {Crystal
  time-reversal symmetry breaking and spontaneous {{Hall}} effect in collinear
  antiferromagnets},\ }\href {https://doi.org/10.1126/sciadv.aaz8809}
  {\bibfield  {journal} {\bibinfo  {journal} {Science Advances}\ }\textbf
  {\bibinfo {volume} {6}},\ \bibinfo {pages} {eaaz8809} (\bibinfo {year}
  {2020})}\BibitemShut {NoStop}%
\bibitem [{\citenamefont {Berlijn}\ \emph {et~al.}(2017)\citenamefont
  {Berlijn}, \citenamefont {Snijders}, \citenamefont {Delaire}, \citenamefont
  {Zhou}, \citenamefont {Maier}, \citenamefont {Cao}, \citenamefont {Chi},
  \citenamefont {Matsuda}, \citenamefont {Wang}, \citenamefont {Koehler},
  \citenamefont {Kent},\ and\ \citenamefont
  {Weitering}}]{berlijnItinerantAntiferromagnetismMathrm2017}%
  \BibitemOpen
  \bibfield  {author} {\bibinfo {author} {\bibfnamefont {T.}~\bibnamefont
  {Berlijn}}, \bibinfo {author} {\bibfnamefont {P.~C.}\ \bibnamefont
  {Snijders}}, \bibinfo {author} {\bibfnamefont {O.}~\bibnamefont {Delaire}},
  \bibinfo {author} {\bibfnamefont {H.-D.}\ \bibnamefont {Zhou}}, \bibinfo
  {author} {\bibfnamefont {T.~A.}\ \bibnamefont {Maier}}, \bibinfo {author}
  {\bibfnamefont {H.-B.}\ \bibnamefont {Cao}}, \bibinfo {author} {\bibfnamefont
  {S.-X.}\ \bibnamefont {Chi}}, \bibinfo {author} {\bibfnamefont
  {M.}~\bibnamefont {Matsuda}}, \bibinfo {author} {\bibfnamefont
  {Y.}~\bibnamefont {Wang}}, \bibinfo {author} {\bibfnamefont {M.~R.}\
  \bibnamefont {Koehler}}, \bibinfo {author} {\bibfnamefont {P.~R.~C.}\
  \bibnamefont {Kent}},\ and\ \bibinfo {author} {\bibfnamefont {H.~H.}\
  \bibnamefont {Weitering}},\ }\bibfield  {title} {\bibinfo {title} {Itinerant
  {{Antiferromagnetism}} in {{RuO}}\textsubscript{2}},\ }\href
  {https://doi.org/10.1103/PhysRevLett.118.077201} {\bibfield  {journal}
  {\bibinfo  {journal} {Physical Review Letters}\ }\textbf {\bibinfo {volume}
  {118}},\ \bibinfo {pages} {077201} (\bibinfo {year} {2017})}\BibitemShut
  {NoStop}%
\bibitem [{\citenamefont {Zhu}\ \emph {et~al.}(2019)\citenamefont {Zhu},
  \citenamefont {Strempfer}, \citenamefont {Rao}, \citenamefont {Occhialini},
  \citenamefont {Pelliciari}, \citenamefont {Choi}, \citenamefont {Kawaguchi},
  \citenamefont {You}, \citenamefont {Mitchell}, \citenamefont {{Shao-Horn}},\
  and\ \citenamefont {Comin}}]{zhuAnomalousAntiferromagnetismMetallic2019}%
  \BibitemOpen
  \bibfield  {author} {\bibinfo {author} {\bibfnamefont {Z.~H.}\ \bibnamefont
  {Zhu}}, \bibinfo {author} {\bibfnamefont {J.}~\bibnamefont {Strempfer}},
  \bibinfo {author} {\bibfnamefont {R.~R.}\ \bibnamefont {Rao}}, \bibinfo
  {author} {\bibfnamefont {C.~A.}\ \bibnamefont {Occhialini}}, \bibinfo
  {author} {\bibfnamefont {J.}~\bibnamefont {Pelliciari}}, \bibinfo {author}
  {\bibfnamefont {Y.}~\bibnamefont {Choi}}, \bibinfo {author} {\bibfnamefont
  {T.}~\bibnamefont {Kawaguchi}}, \bibinfo {author} {\bibfnamefont
  {H.}~\bibnamefont {You}}, \bibinfo {author} {\bibfnamefont {J.~F.}\
  \bibnamefont {Mitchell}}, \bibinfo {author} {\bibfnamefont {Y.}~\bibnamefont
  {{Shao-Horn}}},\ and\ \bibinfo {author} {\bibfnamefont {R.}~\bibnamefont
  {Comin}},\ }\bibfield  {title} {\bibinfo {title} {Anomalous
  {{Antiferromagnetism}} in {{Metallic}} {{RuO}}\textsubscript{2}
  {{Determined}} by {{Resonant X-ray Scattering}}},\ }\href
  {https://doi.org/10.1103/PhysRevLett.122.017202} {\bibfield  {journal}
  {\bibinfo  {journal} {Physical Review Letters}\ }\textbf {\bibinfo {volume}
  {122}},\ \bibinfo {pages} {017202} (\bibinfo {year} {2019})}\BibitemShut
  {NoStop}%
\bibitem [{\citenamefont {Bai}\ \emph {et~al.}(2022)\citenamefont {Bai},
  \citenamefont {Han}, \citenamefont {Feng}, \citenamefont {Zhou},
  \citenamefont {Su}, \citenamefont {Wang}, \citenamefont {Liao}, \citenamefont
  {Zhu}, \citenamefont {Chen}, \citenamefont {Pan}, \citenamefont {Fan},\ and\
  \citenamefont {Song}}]{baiObservationSpinSplitting2022}%
  \BibitemOpen
  \bibfield  {author} {\bibinfo {author} {\bibfnamefont {H.}~\bibnamefont
  {Bai}}, \bibinfo {author} {\bibfnamefont {L.}~\bibnamefont {Han}}, \bibinfo
  {author} {\bibfnamefont {X.~Y.}\ \bibnamefont {Feng}}, \bibinfo {author}
  {\bibfnamefont {Y.~J.}\ \bibnamefont {Zhou}}, \bibinfo {author}
  {\bibfnamefont {R.~X.}\ \bibnamefont {Su}}, \bibinfo {author} {\bibfnamefont
  {Q.}~\bibnamefont {Wang}}, \bibinfo {author} {\bibfnamefont {L.~Y.}\
  \bibnamefont {Liao}}, \bibinfo {author} {\bibfnamefont {W.~X.}\ \bibnamefont
  {Zhu}}, \bibinfo {author} {\bibfnamefont {X.~Z.}\ \bibnamefont {Chen}},
  \bibinfo {author} {\bibfnamefont {F.}~\bibnamefont {Pan}}, \bibinfo {author}
  {\bibfnamefont {X.~L.}\ \bibnamefont {Fan}},\ and\ \bibinfo {author}
  {\bibfnamefont {C.}~\bibnamefont {Song}},\ }\bibfield  {title} {\bibinfo
  {title} {Observation of {{Spin Splitting Torque}} in a {{Collinear
  Antiferromagnet RuO}}\textsubscript{2}},\ }\href
  {https://doi.org/10.1103/PhysRevLett.128.197202} {\bibfield  {journal}
  {\bibinfo  {journal} {Physical Review Letters}\ }\textbf {\bibinfo {volume}
  {128}},\ \bibinfo {pages} {197202} (\bibinfo {year} {2022})}\BibitemShut
  {NoStop}%
\bibitem [{\citenamefont {Bose}\ \emph {et~al.}(2022)\citenamefont {Bose},
  \citenamefont {Schreiber}, \citenamefont {Jain}, \citenamefont {Shao},
  \citenamefont {Nair}, \citenamefont {Sun}, \citenamefont {Zhang},
  \citenamefont {Muller}, \citenamefont {Tsymbal}, \citenamefont {Schlom},\
  and\ \citenamefont {Ralph}}]{boseTiltedSpinCurrent2022}%
  \BibitemOpen
  \bibfield  {author} {\bibinfo {author} {\bibfnamefont {A.}~\bibnamefont
  {Bose}}, \bibinfo {author} {\bibfnamefont {N.~J.}\ \bibnamefont {Schreiber}},
  \bibinfo {author} {\bibfnamefont {R.}~\bibnamefont {Jain}}, \bibinfo {author}
  {\bibfnamefont {D.-F.}\ \bibnamefont {Shao}}, \bibinfo {author}
  {\bibfnamefont {H.~P.}\ \bibnamefont {Nair}}, \bibinfo {author}
  {\bibfnamefont {J.}~\bibnamefont {Sun}}, \bibinfo {author} {\bibfnamefont
  {X.~S.}\ \bibnamefont {Zhang}}, \bibinfo {author} {\bibfnamefont {D.~A.}\
  \bibnamefont {Muller}}, \bibinfo {author} {\bibfnamefont {E.~Y.}\
  \bibnamefont {Tsymbal}}, \bibinfo {author} {\bibfnamefont {D.~G.}\
  \bibnamefont {Schlom}},\ and\ \bibinfo {author} {\bibfnamefont {D.~C.}\
  \bibnamefont {Ralph}},\ }\bibfield  {title} {\bibinfo {title} {Tilted spin
  current generated by the collinear antiferromagnet ruthenium dioxide},\
  }\href {https://doi.org/10.1038/s41928-022-00744-8} {\bibfield  {journal}
  {\bibinfo  {journal} {Nature Electronics}\ }\textbf {\bibinfo {volume} {5}},\
  \bibinfo {pages} {267} (\bibinfo {year} {2022})}\BibitemShut {NoStop}%
\bibitem [{\citenamefont {Fedchenko}\ \emph {et~al.}(2023)\citenamefont
  {Fedchenko}, \citenamefont {Minar}, \citenamefont {Akashdeep}, \citenamefont
  {D'Souza}, \citenamefont {Vasilyev}, \citenamefont {Tkach}, \citenamefont
  {Odenbreit}, \citenamefont {Nguyen}, \citenamefont {Kutnyakhov},
  \citenamefont {Wind}, \citenamefont {Wenthaus}, \citenamefont {Scholz},
  \citenamefont {Rossnagel}, \citenamefont {Hoesch}, \citenamefont
  {Aeschlimann}, \citenamefont {Stadtmueller}, \citenamefont {Klaeui},
  \citenamefont {Schoenhense}, \citenamefont {Jakob}, \citenamefont
  {Jungwirth}, \citenamefont {Smejkal}, \citenamefont {Sinova},\ and\
  \citenamefont {Elmers}}]{fedchenkoObservationTimereversalSymmetry2023}%
  \BibitemOpen
  \bibfield  {author} {\bibinfo {author} {\bibfnamefont {O.}~\bibnamefont
  {Fedchenko}}, \bibinfo {author} {\bibfnamefont {J.}~\bibnamefont {Minar}},
  \bibinfo {author} {\bibfnamefont {A.}~\bibnamefont {Akashdeep}}, \bibinfo
  {author} {\bibfnamefont {S.~W.}\ \bibnamefont {D'Souza}}, \bibinfo {author}
  {\bibfnamefont {D.}~\bibnamefont {Vasilyev}}, \bibinfo {author}
  {\bibfnamefont {O.}~\bibnamefont {Tkach}}, \bibinfo {author} {\bibfnamefont
  {L.}~\bibnamefont {Odenbreit}}, \bibinfo {author} {\bibfnamefont {Q.~L.}\
  \bibnamefont {Nguyen}}, \bibinfo {author} {\bibfnamefont {D.}~\bibnamefont
  {Kutnyakhov}}, \bibinfo {author} {\bibfnamefont {N.}~\bibnamefont {Wind}},
  \bibinfo {author} {\bibfnamefont {L.}~\bibnamefont {Wenthaus}}, \bibinfo
  {author} {\bibfnamefont {M.}~\bibnamefont {Scholz}}, \bibinfo {author}
  {\bibfnamefont {K.}~\bibnamefont {Rossnagel}}, \bibinfo {author}
  {\bibfnamefont {M.}~\bibnamefont {Hoesch}}, \bibinfo {author} {\bibfnamefont
  {M.}~\bibnamefont {Aeschlimann}}, \bibinfo {author} {\bibfnamefont
  {B.}~\bibnamefont {Stadtmueller}}, \bibinfo {author} {\bibfnamefont
  {M.}~\bibnamefont {Klaeui}}, \bibinfo {author} {\bibfnamefont
  {G.}~\bibnamefont {Schoenhense}}, \bibinfo {author} {\bibfnamefont
  {G.}~\bibnamefont {Jakob}}, \bibinfo {author} {\bibfnamefont
  {T.}~\bibnamefont {Jungwirth}}, \bibinfo {author} {\bibfnamefont
  {L.}~\bibnamefont {Smejkal}}, \bibinfo {author} {\bibfnamefont
  {J.}~\bibnamefont {Sinova}},\ and\ \bibinfo {author} {\bibfnamefont {H.~J.}\
  \bibnamefont {Elmers}},\ }\href {https://doi.org/10.48550/arXiv.2306.02170}
  {\bibinfo {title} {Observation of time-reversal symmetry breaking in the band
  structure of altermagnetic {{RuO}}\textsubscript{2}}} (\bibinfo {year}
  {2023}),\ \Eprint {https://arxiv.org/abs/2306.02170} {arxiv:2306.02170
  [cond-mat]} \BibitemShut {NoStop}%
\bibitem [{\citenamefont {Feng}\ \emph {et~al.}(2022)\citenamefont {Feng},
  \citenamefont {Zhou}, \citenamefont {Smejkal}, \citenamefont {Wu},
  \citenamefont {Zhu}, \citenamefont {Guo}, \citenamefont
  {{Gonzalez-Hernandez}}, \citenamefont {Wang}, \citenamefont {Yan},
  \citenamefont {Qin}, \citenamefont {Zhang}, \citenamefont {Wu}, \citenamefont
  {Chen}, \citenamefont {Meng}, \citenamefont {Liu}, \citenamefont {Xia},
  \citenamefont {Sinova}, \citenamefont {Jungwirth},\ and\ \citenamefont
  {Liu}}]{fengAnomalousHallEffect2022}%
  \BibitemOpen
  \bibfield  {author} {\bibinfo {author} {\bibfnamefont {Z.}~\bibnamefont
  {Feng}}, \bibinfo {author} {\bibfnamefont {X.}~\bibnamefont {Zhou}}, \bibinfo
  {author} {\bibfnamefont {L.}~\bibnamefont {Smejkal}}, \bibinfo {author}
  {\bibfnamefont {L.}~\bibnamefont {Wu}}, \bibinfo {author} {\bibfnamefont
  {Z.}~\bibnamefont {Zhu}}, \bibinfo {author} {\bibfnamefont {H.}~\bibnamefont
  {Guo}}, \bibinfo {author} {\bibfnamefont {R.}~\bibnamefont
  {{Gonzalez-Hernandez}}}, \bibinfo {author} {\bibfnamefont {X.}~\bibnamefont
  {Wang}}, \bibinfo {author} {\bibfnamefont {H.}~\bibnamefont {Yan}}, \bibinfo
  {author} {\bibfnamefont {P.}~\bibnamefont {Qin}}, \bibinfo {author}
  {\bibfnamefont {X.}~\bibnamefont {Zhang}}, \bibinfo {author} {\bibfnamefont
  {H.}~\bibnamefont {Wu}}, \bibinfo {author} {\bibfnamefont {H.}~\bibnamefont
  {Chen}}, \bibinfo {author} {\bibfnamefont {Z.}~\bibnamefont {Meng}}, \bibinfo
  {author} {\bibfnamefont {L.}~\bibnamefont {Liu}}, \bibinfo {author}
  {\bibfnamefont {Z.}~\bibnamefont {Xia}}, \bibinfo {author} {\bibfnamefont
  {J.}~\bibnamefont {Sinova}}, \bibinfo {author} {\bibfnamefont
  {T.}~\bibnamefont {Jungwirth}},\ and\ \bibinfo {author} {\bibfnamefont
  {Z.}~\bibnamefont {Liu}},\ }\bibfield  {title} {\bibinfo {title} {An
  anomalous {{Hall}} effect in altermagnetic ruthenium dioxide},\ }\href
  {https://doi.org/10.1038/s41928-022-00866-z} {\bibfield  {journal} {\bibinfo
  {journal} {Nature Electronics}\ }\textbf {\bibinfo {volume} {5}},\ \bibinfo
  {pages} {735} (\bibinfo {year} {2022})}\BibitemShut {NoStop}%
\bibitem [{\citenamefont {Lin}\ \emph {et~al.}(2024)\citenamefont {Lin},
  \citenamefont {Chen}, \citenamefont {Lu}, \citenamefont {Liang},
  \citenamefont {Feng}, \citenamefont {Yamagami}, \citenamefont {Osiecki},
  \citenamefont {Leandersson}, \citenamefont {Thiagarajan}, \citenamefont
  {Liu}, \citenamefont {Felser},\ and\ \citenamefont
  {Ma}}]{linObservationGiantSpin2024}%
  \BibitemOpen
  \bibfield  {author} {\bibinfo {author} {\bibfnamefont {Z.}~\bibnamefont
  {Lin}}, \bibinfo {author} {\bibfnamefont {D.}~\bibnamefont {Chen}}, \bibinfo
  {author} {\bibfnamefont {W.}~\bibnamefont {Lu}}, \bibinfo {author}
  {\bibfnamefont {X.}~\bibnamefont {Liang}}, \bibinfo {author} {\bibfnamefont
  {S.}~\bibnamefont {Feng}}, \bibinfo {author} {\bibfnamefont {K.}~\bibnamefont
  {Yamagami}}, \bibinfo {author} {\bibfnamefont {J.}~\bibnamefont {Osiecki}},
  \bibinfo {author} {\bibfnamefont {M.}~\bibnamefont {Leandersson}}, \bibinfo
  {author} {\bibfnamefont {B.}~\bibnamefont {Thiagarajan}}, \bibinfo {author}
  {\bibfnamefont {J.}~\bibnamefont {Liu}}, \bibinfo {author} {\bibfnamefont
  {C.}~\bibnamefont {Felser}},\ and\ \bibinfo {author} {\bibfnamefont
  {J.}~\bibnamefont {Ma}},\ }\href {https://doi.org/10.48550/arXiv.2402.04995}
  {\bibinfo {title} {Observation of {{Giant Spin Splitting}} and d-wave {{Spin
  Texture}} in {{Room Temperature Altermagnet RuO}}\textsubscript{2}}}
  (\bibinfo {year} {2024}),\ \Eprint {https://arxiv.org/abs/2402.04995}
  {arxiv:2402.04995 [cond-mat]} \BibitemShut {NoStop}%
\bibitem [{\citenamefont {Lovesey}\ \emph {et~al.}(2022)\citenamefont
  {Lovesey}, \citenamefont {Khalyavin},\ and\ \citenamefont {{van der
  Laan}}}]{loveseyMagneticPropertiesMathrm2022}%
  \BibitemOpen
  \bibfield  {author} {\bibinfo {author} {\bibfnamefont {S.~W.}\ \bibnamefont
  {Lovesey}}, \bibinfo {author} {\bibfnamefont {D.~D.}\ \bibnamefont
  {Khalyavin}},\ and\ \bibinfo {author} {\bibfnamefont {G.}~\bibnamefont {{van
  der Laan}}},\ }\bibfield  {title} {\bibinfo {title} {Magnetic properties of
  {{RuO}}\textsubscript{2} and charge-magnetic interference in {{Bragg}}
  diffraction of circularly polarized x-rays},\ }\href
  {https://doi.org/10.1103/PhysRevB.105.014403} {\bibfield  {journal} {\bibinfo
   {journal} {Physical Review B}\ }\textbf {\bibinfo {volume} {105}},\ \bibinfo
  {pages} {014403} (\bibinfo {year} {2022})}\BibitemShut {NoStop}%
\bibitem [{\citenamefont {Smolyanyuk}\ \emph {et~al.}(2024)\citenamefont
  {Smolyanyuk}, \citenamefont {Mazin}, \citenamefont {{Garcia-Gassull}},\ and\
  \citenamefont {Valent{\'i}}}]{smolyanyukFragilityMagneticOrder2024}%
  \BibitemOpen
  \bibfield  {author} {\bibinfo {author} {\bibfnamefont {A.}~\bibnamefont
  {Smolyanyuk}}, \bibinfo {author} {\bibfnamefont {I.~I.}\ \bibnamefont
  {Mazin}}, \bibinfo {author} {\bibfnamefont {L.}~\bibnamefont
  {{Garcia-Gassull}}},\ and\ \bibinfo {author} {\bibfnamefont {R.}~\bibnamefont
  {Valent{\'i}}},\ }\bibfield  {title} {\bibinfo {title} {Fragility of the
  magnetic order in the prototypical altermagnet {{RuO}}\textsubscript{2}},\
  }\href {https://doi.org/10.1103/PhysRevB.109.134424} {\bibfield  {journal}
  {\bibinfo  {journal} {Physical Review B}\ }\textbf {\bibinfo {volume}
  {109}},\ \bibinfo {pages} {134424} (\bibinfo {year} {2024})}\BibitemShut
  {NoStop}%
\bibitem [{\citenamefont {Hiraishi}\ \emph {et~al.}(2024)\citenamefont
  {Hiraishi}, \citenamefont {Okabe}, \citenamefont {Koda}, \citenamefont
  {Kadono}, \citenamefont {Muroi}, \citenamefont {Hirai},\ and\ \citenamefont
  {Hiroi}}]{hiraishiNonmagneticGroundState2024}%
  \BibitemOpen
  \bibfield  {author} {\bibinfo {author} {\bibfnamefont {M.}~\bibnamefont
  {Hiraishi}}, \bibinfo {author} {\bibfnamefont {H.}~\bibnamefont {Okabe}},
  \bibinfo {author} {\bibfnamefont {A.}~\bibnamefont {Koda}}, \bibinfo {author}
  {\bibfnamefont {R.}~\bibnamefont {Kadono}}, \bibinfo {author} {\bibfnamefont
  {T.}~\bibnamefont {Muroi}}, \bibinfo {author} {\bibfnamefont
  {D.}~\bibnamefont {Hirai}},\ and\ \bibinfo {author} {\bibfnamefont
  {Z.}~\bibnamefont {Hiroi}},\ }\bibfield  {title} {\bibinfo {title}
  {Nonmagnetic {{Ground State}} in {{RuO}}\textsubscript{2} {{Revealed}} by
  {{Muon Spin Rotation}}},\ }\href
  {https://doi.org/10.1103/PhysRevLett.132.166702} {\bibfield  {journal}
  {\bibinfo  {journal} {Physical Review Letters}\ }\textbf {\bibinfo {volume}
  {132}},\ \bibinfo {pages} {166702} (\bibinfo {year} {2024})}\BibitemShut
  {NoStop}%
\bibitem [{\citenamefont {Glassford}\ and\ \citenamefont
  {Chelikowsky}(1994)}]{glassfordElectronTransportProperties1994}%
  \BibitemOpen
  \bibfield  {author} {\bibinfo {author} {\bibfnamefont {K.~M.}\ \bibnamefont
  {Glassford}}\ and\ \bibinfo {author} {\bibfnamefont {J.~R.}\ \bibnamefont
  {Chelikowsky}},\ }\bibfield  {title} {\bibinfo {title} {Electron transport
  properties in {{RuO}}\textsubscript{2} rutile},\ }\href
  {https://doi.org/10.1103/PhysRevB.49.7107} {\bibfield  {journal} {\bibinfo
  {journal} {Physical Review B}\ }\textbf {\bibinfo {volume} {49}},\ \bibinfo
  {pages} {7107} (\bibinfo {year} {1994})}\BibitemShut {NoStop}%
\bibitem [{\citenamefont {Huang}\ \emph {et~al.}(1982)\citenamefont {Huang},
  \citenamefont {Park},\ and\ \citenamefont
  {Pollak}}]{huangGrowthCharacterizationRuO21982}%
  \BibitemOpen
  \bibfield  {author} {\bibinfo {author} {\bibfnamefont {Y.}~\bibnamefont
  {Huang}}, \bibinfo {author} {\bibfnamefont {H.}~\bibnamefont {Park}},\ and\
  \bibinfo {author} {\bibfnamefont {F.~H.}\ \bibnamefont {Pollak}},\ }\bibfield
   {title} {\bibinfo {title} {Growth and characterization of
  {{RuO}}\textsubscript{2} single crystals},\ }\href
  {https://doi.org/10.1016/0025-5408(82)90166-0} {\bibfield  {journal}
  {\bibinfo  {journal} {Materials Research Bulletin}\ }\textbf {\bibinfo
  {volume} {17}},\ \bibinfo {pages} {1305} (\bibinfo {year}
  {1982})}\BibitemShut {NoStop}%
\bibitem [{\citenamefont {Rao}\ and\ \citenamefont
  {Iyengar}(1969)}]{raoXrayStudiesThermal1969}%
  \BibitemOpen
  \bibfield  {author} {\bibinfo {author} {\bibfnamefont {K.~V.~K.}\
  \bibnamefont {Rao}}\ and\ \bibinfo {author} {\bibfnamefont {L.}~\bibnamefont
  {Iyengar}},\ }\bibfield  {title} {\bibinfo {title} {X-ray studies on the
  thermal expansion of ruthenium dioxide},\ }\href
  {https://doi.org/10.1107/S0567739469000465} {\bibfield  {journal} {\bibinfo
  {journal} {Acta Crystallographica Section A}\ }\textbf {\bibinfo {volume}
  {25}},\ \bibinfo {pages} {302} (\bibinfo {year} {1969})}\BibitemShut
  {NoStop}%
\bibitem [{\citenamefont {Butler}\ and\ \citenamefont
  {Gillson}(1971)}]{butlerCrystalGrowthElectrical1971}%
  \BibitemOpen
  \bibfield  {author} {\bibinfo {author} {\bibfnamefont {S.}~\bibnamefont
  {Butler}}\ and\ \bibinfo {author} {\bibfnamefont {J.}~\bibnamefont
  {Gillson}},\ }\bibfield  {title} {\bibinfo {title} {Crystal growth,
  electrical resistivity and lattice parameters of {{RuO}}\textsubscript{2} and
  {{IrO}}\textsubscript{2}},\ }\href
  {https://doi.org/10.1016/0025-5408(71)90092-4} {\bibfield  {journal}
  {\bibinfo  {journal} {Materials Research Bulletin}\ }\textbf {\bibinfo
  {volume} {6}},\ \bibinfo {pages} {81} (\bibinfo {year} {1971})}\BibitemShut
  {NoStop}%
\bibitem [{\citenamefont {Uchida}\ \emph {et~al.}(2020)\citenamefont {Uchida},
  \citenamefont {Nomoto}, \citenamefont {Musashi}, \citenamefont {Arita},\ and\
  \citenamefont {Kawasaki}}]{uchidaSuperconductivityUniquelyStrained2020}%
  \BibitemOpen
  \bibfield  {author} {\bibinfo {author} {\bibfnamefont {M.}~\bibnamefont
  {Uchida}}, \bibinfo {author} {\bibfnamefont {T.}~\bibnamefont {Nomoto}},
  \bibinfo {author} {\bibfnamefont {M.}~\bibnamefont {Musashi}}, \bibinfo
  {author} {\bibfnamefont {R.}~\bibnamefont {Arita}},\ and\ \bibinfo {author}
  {\bibfnamefont {M.}~\bibnamefont {Kawasaki}},\ }\bibfield  {title} {\bibinfo
  {title} {Superconductivity in {{Uniquely Strained}} {{RuO}}\textsubscript{2}
  {{Films}}},\ }\href {https://doi.org/10.1103/PhysRevLett.125.147001}
  {\bibfield  {journal} {\bibinfo  {journal} {Physical Review Letters}\
  }\textbf {\bibinfo {volume} {125}},\ \bibinfo {pages} {147001} (\bibinfo
  {year} {2020})}\BibitemShut {NoStop}%
\bibitem [{\citenamefont {Cui}\ \emph {et~al.}(2023)\citenamefont {Cui},
  \citenamefont {Li}, \citenamefont {Chen}, \citenamefont {Chen}, \citenamefont
  {Wu}, \citenamefont {Pei}, \citenamefont {Wu}, \citenamefont {Xie},
  \citenamefont {Che}, \citenamefont {Qiu}, \citenamefont {Liu}, \citenamefont
  {Yuan},\ and\ \citenamefont {Wu}}]{cuiInplaneHallEffect2023}%
  \BibitemOpen
  \bibfield  {author} {\bibinfo {author} {\bibfnamefont {Y.}~\bibnamefont
  {Cui}}, \bibinfo {author} {\bibfnamefont {Z.}~\bibnamefont {Li}}, \bibinfo
  {author} {\bibfnamefont {H.}~\bibnamefont {Chen}}, \bibinfo {author}
  {\bibfnamefont {Y.}~\bibnamefont {Chen}}, \bibinfo {author} {\bibfnamefont
  {Y.}~\bibnamefont {Wu}}, \bibinfo {author} {\bibfnamefont {K.}~\bibnamefont
  {Pei}}, \bibinfo {author} {\bibfnamefont {T.}~\bibnamefont {Wu}}, \bibinfo
  {author} {\bibfnamefont {N.}~\bibnamefont {Xie}}, \bibinfo {author}
  {\bibfnamefont {R.}~\bibnamefont {Che}}, \bibinfo {author} {\bibfnamefont
  {X.}~\bibnamefont {Qiu}}, \bibinfo {author} {\bibfnamefont {Y.}~\bibnamefont
  {Liu}}, \bibinfo {author} {\bibfnamefont {Z.}~\bibnamefont {Yuan}},\ and\
  \bibinfo {author} {\bibfnamefont {Y.}~\bibnamefont {Wu}},\ }\href
  {https://doi.org/10.48550/arXiv.2308.06651} {\bibinfo {title} {In-plane
  {{Hall}} effect in rutile oxide films induced by the {{Lorentz}} force}}
  (\bibinfo {year} {2023}),\ \Eprint {https://arxiv.org/abs/2308.06651}
  {arxiv:2308.06651 [cond-mat]} \BibitemShut {NoStop}%
\bibitem [{\citenamefont {Lin}\ \emph {et~al.}(2004)\citenamefont {Lin},
  \citenamefont {Huang}, \citenamefont {Lin}, \citenamefont {Lee},
  \citenamefont {Liu}, \citenamefont {Zhang}, \citenamefont {Chen},\ and\
  \citenamefont {Huang}}]{linLowTemperatureElectrical2004}%
  \BibitemOpen
  \bibfield  {author} {\bibinfo {author} {\bibfnamefont {J.~J.}\ \bibnamefont
  {Lin}}, \bibinfo {author} {\bibfnamefont {S.~M.}\ \bibnamefont {Huang}},
  \bibinfo {author} {\bibfnamefont {Y.~H.}\ \bibnamefont {Lin}}, \bibinfo
  {author} {\bibfnamefont {T.~C.}\ \bibnamefont {Lee}}, \bibinfo {author}
  {\bibfnamefont {H.}~\bibnamefont {Liu}}, \bibinfo {author} {\bibfnamefont
  {X.~X.}\ \bibnamefont {Zhang}}, \bibinfo {author} {\bibfnamefont {R.~S.}\
  \bibnamefont {Chen}},\ and\ \bibinfo {author} {\bibfnamefont {Y.~S.}\
  \bibnamefont {Huang}},\ }\bibfield  {title} {\bibinfo {title} {Low
  temperature electrical transport properties of {{RuO}}\textsubscript{2} and
  {{IrO}}{\textsubscript{2}} single crystals},\ }\href
  {https://doi.org/10.1088/0953-8984/16/45/025} {\bibfield  {journal} {\bibinfo
   {journal} {Journal of Physics: Condensed Matter}\ }\textbf {\bibinfo
  {volume} {16}},\ \bibinfo {pages} {8035} (\bibinfo {year}
  {2004})}\BibitemShut {NoStop}%
\bibitem [{\citenamefont {Hartnoll}\ and\ \citenamefont
  {Mackenzie}(2022)}]{hartnollColloquiumPlanckianDissipation2022}%
  \BibitemOpen
  \bibfield  {author} {\bibinfo {author} {\bibfnamefont {S.~A.}\ \bibnamefont
  {Hartnoll}}\ and\ \bibinfo {author} {\bibfnamefont {A.~P.}\ \bibnamefont
  {Mackenzie}},\ }\bibfield  {title} {\bibinfo {title} {Colloquium:
  {{Planckian}} dissipation in metals},\ }\href
  {https://doi.org/10.1103/RevModPhys.94.041002} {\bibfield  {journal}
  {\bibinfo  {journal} {Reviews of Modern Physics}\ }\textbf {\bibinfo {volume}
  {94}},\ \bibinfo {pages} {041002} (\bibinfo {year} {2022})}\BibitemShut
  {NoStop}%
\bibitem [{\citenamefont {Grissonnanche}\ \emph {et~al.}(2021)\citenamefont
  {Grissonnanche}, \citenamefont {Fang}, \citenamefont {Legros}, \citenamefont
  {Verret}, \citenamefont {Lalibert{\'e}}, \citenamefont {Collignon},
  \citenamefont {Zhou}, \citenamefont {Graf}, \citenamefont {Goddard},
  \citenamefont {Taillefer},\ and\ \citenamefont
  {Ramshaw}}]{grissonnancheLinearinTemperatureResistivity2021}%
  \BibitemOpen
  \bibfield  {author} {\bibinfo {author} {\bibfnamefont {G.}~\bibnamefont
  {Grissonnanche}}, \bibinfo {author} {\bibfnamefont {Y.}~\bibnamefont {Fang}},
  \bibinfo {author} {\bibfnamefont {A.}~\bibnamefont {Legros}}, \bibinfo
  {author} {\bibfnamefont {S.}~\bibnamefont {Verret}}, \bibinfo {author}
  {\bibfnamefont {F.}~\bibnamefont {Lalibert{\'e}}}, \bibinfo {author}
  {\bibfnamefont {C.}~\bibnamefont {Collignon}}, \bibinfo {author}
  {\bibfnamefont {J.}~\bibnamefont {Zhou}}, \bibinfo {author} {\bibfnamefont
  {D.}~\bibnamefont {Graf}}, \bibinfo {author} {\bibfnamefont {P.~A.}\
  \bibnamefont {Goddard}}, \bibinfo {author} {\bibfnamefont {L.}~\bibnamefont
  {Taillefer}},\ and\ \bibinfo {author} {\bibfnamefont {B.~J.}\ \bibnamefont
  {Ramshaw}},\ }\bibfield  {title} {\bibinfo {title} {Linear-in temperature
  resistivity from an isotropic {{Planckian}} scattering rate},\ }\href
  {https://doi.org/10.1038/s41586-021-03697-8} {\bibfield  {journal} {\bibinfo
  {journal} {Nature}\ }\textbf {\bibinfo {volume} {595}},\ \bibinfo {pages}
  {667} (\bibinfo {year} {2021})}\BibitemShut {NoStop}%
\bibitem [{\citenamefont {H{\'e}bert}\ \emph {et~al.}(2015)\citenamefont
  {H{\'e}bert}, \citenamefont {Daou},\ and\ \citenamefont
  {Maignan}}]{hebertThermopowerQuadruplePerovskite2015}%
  \BibitemOpen
  \bibfield  {author} {\bibinfo {author} {\bibfnamefont {S.}~\bibnamefont
  {H{\'e}bert}}, \bibinfo {author} {\bibfnamefont {R.}~\bibnamefont {Daou}},\
  and\ \bibinfo {author} {\bibfnamefont {A.}~\bibnamefont {Maignan}},\
  }\bibfield  {title} {\bibinfo {title} {Thermopower in the quadruple
  perovskite ruthenates},\ }\href {https://doi.org/10.1103/PhysRevB.91.045106}
  {\bibfield  {journal} {\bibinfo  {journal} {Physical Review B}\ }\textbf
  {\bibinfo {volume} {91}},\ \bibinfo {pages} {045106} (\bibinfo {year}
  {2015})}\BibitemShut {NoStop}%
\bibitem [{\citenamefont {Pawula}\ \emph {et~al.}(2021)\citenamefont {Pawula},
  \citenamefont {Daou}, \citenamefont {H{\'e}bert},\ and\ \citenamefont
  {Maignan}}]{pawulaThermoelectricPropertiesStandard2021}%
  \BibitemOpen
  \bibfield  {author} {\bibinfo {author} {\bibfnamefont {F.}~\bibnamefont
  {Pawula}}, \bibinfo {author} {\bibfnamefont {R.}~\bibnamefont {Daou}},
  \bibinfo {author} {\bibfnamefont {S.}~\bibnamefont {H{\'e}bert}},\ and\
  \bibinfo {author} {\bibfnamefont {A.}~\bibnamefont {Maignan}},\ }\bibfield
  {title} {\bibinfo {title} {1.1 - {{Thermoelectric}} properties beyond the
  standard {{Boltzmann}} model in oxides: {{A}} focus on the ruthenates},\ }in\
  \href {https://doi.org/10.1016/B978-0-12-818535-3.00028-1} {\emph {\bibinfo
  {booktitle} {Thermoelectric {{Energy Conversion}}}}},\ \bibinfo {series and
  number} {Woodhead {{Publishing Series}} in {{Electronic}} and {{Optical
  Materials}}},\ \bibinfo {editor} {edited by\ \bibinfo {editor} {\bibfnamefont
  {R.}~\bibnamefont {Funahashi}}}\ (\bibinfo  {publisher} {{Woodhead
  Publishing}},\ \bibinfo {year} {2021})\ pp.\ \bibinfo {pages}
  {3--14}\BibitemShut {NoStop}%
\bibitem [{\citenamefont {Mravlje}\ and\ \citenamefont
  {Georges}(2016)}]{mravljeThermopowerEntropyLessons2016a}%
  \BibitemOpen
  \bibfield  {author} {\bibinfo {author} {\bibfnamefont {J.}~\bibnamefont
  {Mravlje}}\ and\ \bibinfo {author} {\bibfnamefont {A.}~\bibnamefont
  {Georges}},\ }\bibfield  {title} {\bibinfo {title} {Thermopower and
  {{Entropy}}: Lessons from {{Sr}}\textsubscript{2}{{RuO}}\textsubscript{4}},\
  }\href {https://doi.org/10.1103/PhysRevLett.117.036401} {\bibfield  {journal}
  {\bibinfo  {journal} {Physical Review Letters}\ }\textbf {\bibinfo {volume}
  {117}},\ \bibinfo {pages} {036401} (\bibinfo {year} {2016})},\ \Eprint
  {https://arxiv.org/abs/1504.03860} {arxiv:1504.03860 [cond-mat]} \BibitemShut
  {NoStop}%
\bibitem [{\citenamefont {Yavorsky}\ \emph {et~al.}(1996)\citenamefont
  {Yavorsky}, \citenamefont {Krasovska}, \citenamefont {Krasovskii},
  \citenamefont {Yaresko},\ and\ \citenamefont
  {Antonov}}]{yavorskyInitioCalculationFermi1996}%
  \BibitemOpen
  \bibfield  {author} {\bibinfo {author} {\bibfnamefont {{\relax
  B.Yu}.}~\bibnamefont {Yavorsky}}, \bibinfo {author} {\bibfnamefont
  {O.}~\bibnamefont {Krasovska}}, \bibinfo {author} {\bibfnamefont
  {E.}~\bibnamefont {Krasovskii}}, \bibinfo {author} {\bibfnamefont
  {A.}~\bibnamefont {Yaresko}},\ and\ \bibinfo {author} {\bibfnamefont
  {V.}~\bibnamefont {Antonov}},\ }\bibfield  {title} {\bibinfo {title} {Ab
  initio calculation of the {{Fermi}} surface of {{RuO}}\textsubscript{2}},\
  }\href {https://doi.org/10.1016/0921-4526(96)00270-0} {\bibfield  {journal}
  {\bibinfo  {journal} {Physica B: Condensed Matter}\ }\textbf {\bibinfo
  {volume} {225}},\ \bibinfo {pages} {243} (\bibinfo {year}
  {1996})}\BibitemShut {NoStop}%
\bibitem [{\citenamefont {Madsen}\ \emph {et~al.}(2018)\citenamefont {Madsen},
  \citenamefont {Carrete},\ and\ \citenamefont
  {Verstraete}}]{madsenBoltzTraP2ProgramInterpolating2018}%
  \BibitemOpen
  \bibfield  {author} {\bibinfo {author} {\bibfnamefont {G.~K.~H.}\
  \bibnamefont {Madsen}}, \bibinfo {author} {\bibfnamefont {J.}~\bibnamefont
  {Carrete}},\ and\ \bibinfo {author} {\bibfnamefont {M.~J.}\ \bibnamefont
  {Verstraete}},\ }\bibfield  {title} {\bibinfo {title} {{{BoltzTraP2}}, a
  program for interpolating band structures and calculating semi-classical
  transport coefficients},\ }\href {https://doi.org/10.1016/j.cpc.2018.05.010}
  {\bibfield  {journal} {\bibinfo  {journal} {Computer Physics Communications}\
  }\textbf {\bibinfo {volume} {231}},\ \bibinfo {pages} {140} (\bibinfo {year}
  {2018})}\BibitemShut {NoStop}%
\bibitem [{\citenamefont {Georges}\ and\ \citenamefont
  {Mravlje}(2021)}]{georgesSkewedNonFermiLiquids2021}%
  \BibitemOpen
  \bibfield  {author} {\bibinfo {author} {\bibfnamefont {A.}~\bibnamefont
  {Georges}}\ and\ \bibinfo {author} {\bibfnamefont {J.}~\bibnamefont
  {Mravlje}},\ }\bibfield  {title} {\bibinfo {title} {Skewed non-{{Fermi}}
  liquids and the {{Seebeck}} effect},\ }\href
  {https://doi.org/10.1103/PhysRevResearch.3.043132} {\bibfield  {journal}
  {\bibinfo  {journal} {Physical Review Research}\ }\textbf {\bibinfo {volume}
  {3}},\ \bibinfo {pages} {043132} (\bibinfo {year} {2021})}\BibitemShut
  {NoStop}%
\bibitem [{\citenamefont {Gourgout}\ \emph {et~al.}(2022)\citenamefont
  {Gourgout}, \citenamefont {Grissonnanche}, \citenamefont {Lalibert{\'e}},
  \citenamefont {Ataei}, \citenamefont {Chen}, \citenamefont {Verret},
  \citenamefont {Zhou}, \citenamefont {Mravlje}, \citenamefont {Georges},
  \citenamefont {{Doiron-Leyraud}},\ and\ \citenamefont
  {Taillefer}}]{gourgoutSeebeckCoefficientCuprate2022}%
  \BibitemOpen
  \bibfield  {author} {\bibinfo {author} {\bibfnamefont {A.}~\bibnamefont
  {Gourgout}}, \bibinfo {author} {\bibfnamefont {G.}~\bibnamefont
  {Grissonnanche}}, \bibinfo {author} {\bibfnamefont {F.}~\bibnamefont
  {Lalibert{\'e}}}, \bibinfo {author} {\bibfnamefont {A.}~\bibnamefont
  {Ataei}}, \bibinfo {author} {\bibfnamefont {L.}~\bibnamefont {Chen}},
  \bibinfo {author} {\bibfnamefont {S.}~\bibnamefont {Verret}}, \bibinfo
  {author} {\bibfnamefont {J.-S.}\ \bibnamefont {Zhou}}, \bibinfo {author}
  {\bibfnamefont {J.}~\bibnamefont {Mravlje}}, \bibinfo {author} {\bibfnamefont
  {A.}~\bibnamefont {Georges}}, \bibinfo {author} {\bibfnamefont
  {N.}~\bibnamefont {{Doiron-Leyraud}}},\ and\ \bibinfo {author} {\bibfnamefont
  {L.}~\bibnamefont {Taillefer}},\ }\bibfield  {title} {\bibinfo {title}
  {Seebeck {{Coefficient}} in a {{Cuprate Superconductor}}: {{Particle-Hole
  Asymmetry}} in the {{Strange Metal Phase}} and {{Fermi Surface
  Transformation}} in the {{Pseudogap Phase}}},\ }\href
  {https://doi.org/10.1103/PhysRevX.12.011037} {\bibfield  {journal} {\bibinfo
  {journal} {Physical Review X}\ }\textbf {\bibinfo {volume} {12}},\ \bibinfo
  {pages} {011037} (\bibinfo {year} {2022})}\BibitemShut {NoStop}%
\bibitem [{\citenamefont
  {Marcus}(1968)}]{marcusMeasurementMagnetoresistanceTransition1968}%
  \BibitemOpen
  \bibfield  {author} {\bibinfo {author} {\bibfnamefont {S.~M.}\ \bibnamefont
  {Marcus}},\ }\bibfield  {title} {\bibinfo {title} {Measurement of the
  magnetoresistance in the transition metal oxide {{RuO}}\textsubscript{2}},\
  }\href {https://doi.org/10.1016/0375-9601(68)90194-1} {\bibfield  {journal}
  {\bibinfo  {journal} {Physics Letters A}\ }\textbf {\bibinfo {volume} {28}},\
  \bibinfo {pages} {191} (\bibinfo {year} {1968})}\BibitemShut {NoStop}%
\bibitem [{\citenamefont
  {Mattheiss}(1976)}]{mattheissElectronicStructureRuO1976}%
  \BibitemOpen
  \bibfield  {author} {\bibinfo {author} {\bibfnamefont {L.~F.}\ \bibnamefont
  {Mattheiss}},\ }\bibfield  {title} {\bibinfo {title} {Electronic structure of
  {{RuO}}\textsubscript{2}, {{OsO}}\textsubscript{2}, and
  {{IrO}}\textsubscript{2}},\ }\href {https://doi.org/10.1103/PhysRevB.13.2433}
  {\bibfield  {journal} {\bibinfo  {journal} {Physical Review B}\ }\textbf
  {\bibinfo {volume} {13}},\ \bibinfo {pages} {2433} (\bibinfo {year}
  {1976})}\BibitemShut {NoStop}%
\bibitem [{\citenamefont {Graebner}\ \emph {et~al.}(1976)\citenamefont
  {Graebner}, \citenamefont {Greiner},\ and\ \citenamefont
  {Ryden}}]{graebnerMagnetothermalOscillationsRuO1976}%
  \BibitemOpen
  \bibfield  {author} {\bibinfo {author} {\bibfnamefont {J.~E.}\ \bibnamefont
  {Graebner}}, \bibinfo {author} {\bibfnamefont {E.~S.}\ \bibnamefont
  {Greiner}},\ and\ \bibinfo {author} {\bibfnamefont {W.~D.}\ \bibnamefont
  {Ryden}},\ }\bibfield  {title} {\bibinfo {title} {Magnetothermal oscillations
  in {{RuO}}\textsubscript{2}, {{OsO}}\textsubscript{2}, and
  {{IrO}}\textsubscript{2}},\ }\href {https://doi.org/10.1103/PhysRevB.13.2426}
  {\bibfield  {journal} {\bibinfo  {journal} {Physical Review B}\ }\textbf
  {\bibinfo {volume} {13}},\ \bibinfo {pages} {2426} (\bibinfo {year}
  {1976})}\BibitemShut {NoStop}%
\bibitem [{\citenamefont
  {Steeves}(2011)}]{steevesELECTRONICTRANSPORTPROPERTIES}%
  \BibitemOpen
  \bibfield  {author} {\bibinfo {author} {\bibfnamefont {M.~M.}\ \bibnamefont
  {Steeves}},\ }\emph {\bibinfo {title} {Electronic Transport Properties of
  Ruthenium and Ruthenium Dioxide Thin Films}},\ \href@noop {} {\bibinfo {type}
  {Phd thesis}},\ \bibinfo  {school} {University of Maine} (\bibinfo {year}
  {2011}),\ \bibinfo {note} {available at
  \url{https://digitalcommons.library.umaine.edu/etd/262}}\BibitemShut
  {NoStop}%
\bibitem [{\citenamefont {Tschirner}\ \emph {et~al.}(2023)\citenamefont
  {Tschirner}, \citenamefont {Ke{\ss}ler}, \citenamefont {Betancourt},
  \citenamefont {Kotte}, \citenamefont {Kriegner}, \citenamefont {Buechner},
  \citenamefont {Dufouleur}, \citenamefont {Kamp}, \citenamefont {Jovic},
  \citenamefont {Smejkal}, \citenamefont {Sinova}, \citenamefont {Claessen},
  \citenamefont {Jungwirth}, \citenamefont {Moser}, \citenamefont {Reichlova},\
  and\ \citenamefont {Veyrat}}]{tschirnerSaturationAnomalousHall2023}%
  \BibitemOpen
  \bibfield  {author} {\bibinfo {author} {\bibfnamefont {T.}~\bibnamefont
  {Tschirner}}, \bibinfo {author} {\bibfnamefont {P.}~\bibnamefont
  {Ke{\ss}ler}}, \bibinfo {author} {\bibfnamefont {R.~D.~G.}\ \bibnamefont
  {Betancourt}}, \bibinfo {author} {\bibfnamefont {T.}~\bibnamefont {Kotte}},
  \bibinfo {author} {\bibfnamefont {D.}~\bibnamefont {Kriegner}}, \bibinfo
  {author} {\bibfnamefont {B.}~\bibnamefont {Buechner}}, \bibinfo {author}
  {\bibfnamefont {J.}~\bibnamefont {Dufouleur}}, \bibinfo {author}
  {\bibfnamefont {M.}~\bibnamefont {Kamp}}, \bibinfo {author} {\bibfnamefont
  {V.}~\bibnamefont {Jovic}}, \bibinfo {author} {\bibfnamefont
  {L.}~\bibnamefont {Smejkal}}, \bibinfo {author} {\bibfnamefont
  {J.}~\bibnamefont {Sinova}}, \bibinfo {author} {\bibfnamefont
  {R.}~\bibnamefont {Claessen}}, \bibinfo {author} {\bibfnamefont
  {T.}~\bibnamefont {Jungwirth}}, \bibinfo {author} {\bibfnamefont
  {S.}~\bibnamefont {Moser}}, \bibinfo {author} {\bibfnamefont
  {H.}~\bibnamefont {Reichlova}},\ and\ \bibinfo {author} {\bibfnamefont
  {L.}~\bibnamefont {Veyrat}},\ }\bibfield  {title} {\bibinfo {title}
  {Saturation of the anomalous {{Hall}} effect at high magnetic fields in
  altermagnetic {{RuO}}\textsubscript{2}}\ }\href
  {https://doi.org/10.48550/arXiv.2309.00568} {10.48550/arXiv.2309.00568}
  (\bibinfo {year} {2023}),\ \Eprint {https://arxiv.org/abs/2309.00568}
  {arxiv:2309.00568 [cond-mat]} \BibitemShut {NoStop}%
\bibitem [{\citenamefont {Sasabe}\ \emph {et~al.}(2023)\citenamefont {Sasabe},
  \citenamefont {Mizumaki}, \citenamefont {Uozumi},\ and\ \citenamefont
  {Yamasaki}}]{sasabeFerroicOrderAnisotropic2023}%
  \BibitemOpen
  \bibfield  {author} {\bibinfo {author} {\bibfnamefont {N.}~\bibnamefont
  {Sasabe}}, \bibinfo {author} {\bibfnamefont {M.}~\bibnamefont {Mizumaki}},
  \bibinfo {author} {\bibfnamefont {T.}~\bibnamefont {Uozumi}},\ and\ \bibinfo
  {author} {\bibfnamefont {Y.}~\bibnamefont {Yamasaki}},\ }\bibfield  {title}
  {\bibinfo {title} {Ferroic {{Order}} for {{Anisotropic Magnetic Dipole Term}}
  in {{Collinear Antiferromagnets}} of (t\textsubscript{2g})\textsuperscript{4}
  {{System}}},\ }\href {https://doi.org/10.1103/PhysRevLett.131.216501}
  {\bibfield  {journal} {\bibinfo  {journal} {Physical Review Letters}\
  }\textbf {\bibinfo {volume} {131}},\ \bibinfo {pages} {216501} (\bibinfo
  {year} {2023})}\BibitemShut {NoStop}%
\end{thebibliography}
%

\end{document}